\documentstyle[12pt,fleqn]{article}
\def\soc{{\rm C}_{60}}
\def\rug{{\rm C}_{70}}

\def\beeq{\begin{equation}}
\def\eneq{\end{equation}}
\def\beeqa{\begin{eqnarray}}
\def\eneqa{\end{eqnarray}}

\addtocounter{section}{-1}
\begin{document}

\begin{center}

\vspace{2in}

{\large {\bf{Charge transfer excitons 
in C$_{\bf 60}$-dimers and polymers\\
} } }

\vspace{1cm}

{\rm Kikuo Harigaya\footnote[1]{E-mail address: 
harigaya@etl.go.jp; URL: http://www.etl.go.jp/People/harigaya/}}\\

\vspace{1cm}

{\sl Physical Science Division,
Electrotechnical Laboratory,\\ 
Umezono 1-1-4, Tsukuba, Ibaraki 305, Japan}

\vspace{1cm}

(Received~~~~~~~~~~~~~~~~~~~~~~~~~~~~~~~~~~~)
\end{center}

\vspace{1cm}

\noindent
{\bf ABSTRACT}\\
Charge-transfer (CT) exciton effects are investigated 
for the optical absorption spectra of crosslinked $\soc$ 
systems by using the intermediate exciton theory.  We 
consider the $\soc$-dimers, and the two (and three) molecule 
systems of the $\soc$-polymers.  We use a tight-binding model 
with long-range Coulomb interactions among electrons, and 
the model is treated by the Hartree-Fock approximation 
followed by the single-excitation configuration interaction 
method.  We discuss the variations in the optical spectra 
by changing the conjugation parameter between molecules.  
We find that the total CT-component increases in smaller 
conjugations, and saturates at the intermediate conjugations. 
It decreases in the large conjugations.   We also find that 
the CT-components of the doped systems are smaller than 
those of the neutral systems, indicating that the electron-hole 
distance becomes shorter in the doped $\soc$-polymers.



\pagebreak

\section{INTRODUCTION}

Recently, it has been revealed that the $\soc$ molecules 
polymerize under UV irradiation [1], and also by applying 
high pressures [2-4].  In these polymerized forms of 
$\soc$, the number of electrons does not change from that 
of $\soc$ molecules, and the systems are kept neutral.  
And further, the linear (one-dimensional) $\soc$-polymers 
have been synthesized in alkali-metal doped $\soc$ crystals: 
$A\soc$ ($A=$K, Rb, Cs) [5-8].  One electron per one $\soc$ 
is donated in the polymer chain in these compounds.  The 
lattice structures are shown in Fig. 1.  The $\soc$ molecules 
are arrayed in a linear chain.  Between the neighboring $\soc$ 
molecules, there are four membered rings which are formed 
by the [2+2] cycloaddition.

In the previous papers [9-10], we have studied the electronic 
structures of one dimensional $\soc$-polymers by using an 
interacting electron-phonon model.  We have introduced a 
phenomenological parameter which represents the strength of 
conjugations of electrons in the polymer chain direction.  
When the conjugations are stronger, the hopping integrals 
between $\soc$ molecules become larger.  We have found that 
the level crossing between the highest occupied state and 
the lowest unoccupied state occurs when the conjugations
increase in the neutral polymer [9].  We have discussed that 
this fact might give rise to a reentrant transition between 
insulators and a metal while a high pressure is applied in 
order to change conjugations between molecules.  The electron
doping effects have been further studied [10].   We have 
found that the $\soc$-polymer doped with one electron per 
one molecule is always a metal, and the polymer doped with 
two electrons per one $\soc$ changes from an indirect-gap 
insulator to the direct-gap insulator, as the conjugations 
become stronger.

A lot of optical experiments on the neutral $\soc$ thin films
and the alkali-metal doped crystals have been performed
(see reviews, refs. [11] and [12], for example).  Along with the
experimental develops, we have theoretically analyzed the 
optical spectra of $\soc$ and $\rug$ molecules [13] and 
solids [14,15].  The $\soc$ and $\rug$ systems maximally 
doped with alkali metals have been also studied [16].
We have searched for parameter sets in order to explain the 
photoexcitation energies, and relative oscillator strengths, 
which have been measured in experiments of $\soc$ and $\rug$ 
systems.  We now expect that our experiences on exciton 
effects in fullerene systems can be well applied to the 
$\soc$-polymers, too.

In this paper, we shall study optical excitation properties
of the polymerized $\soc$ systems by using the intermediate 
exciton theory used in the above literatures.  We consider 
the $\soc$-dimers and the two-molecule system of the 
$\soc$-polymers (and three-molecule system in the Appendix), 
as the examples of the crosslinked $\soc$.  
We use a tight-binding Hamiltonian with long-range Coulomb 
interactions among electrons.  The model has been used in 
the discussion of the antiferromagnetism in the previous 
paper [17].  We shall perform Hartree-Fock approximation 
and take into account of electron-hole excitations by the
single-excitation configuration interaction method.  This 
is a sufficient method when we discuss linear excitation 
properties.  An interesting problem, special to the polymerized 
systems, is the possibility of the crossover among the Frenkel
(molecular) excitons and Wannier [charge-transfer (CT)] excitons.
If interactions are not present between molecules, the exciton
is localized on a molecule.  As the interactions turn on,
some of the optical excitations might have amplitudes 
over several molecules, and thus the excitations become
like CT-excitons.  The main purpose of this paper is to 
look at the development of CT-exciton features in the
optical absorption spectra of $\soc$-dimers and polymers,
when the electronic system is neutral and is doped with
one electron per one $\soc$.

Our main conclusions are:  (1) The CT-exciton component first 
increases in smaller conjugations, i.e., in weak intermolecular
interactions, and saturates at the intermediate conjugations.
The component begins to decrease in the large conjugations.
This fact is commonly seen in the $\soc$-dimers and polymers. 
(2)  The CT-components of the one-electron doped systems 
per $\soc$ are smaller than those of the neutral systems.
The electron-hole separations become shorter 
upon doping in polymerized $\soc$ systems.

This paper is organized as follows.  In the next section,
we present our model and describe procedures of calculations.
Section 3 is devoted to the results of $\soc$-dimers,
and sec. 4 is for the results of $\soc$-polymers.
We summarize the paper in sec. 5.  The three-molecule system
is compared with the two-molecule system in the Appendix.

\section{MODEL}

We use the following tight-binding model with Coulomb 
interactions among electrons.
\beeqa
H &=& H_{\rm pol} + H_{\rm int}, \\
H_{\rm pol} &=&  - at \sum_{l,\sigma} 
\sum_{\langle i,j \rangle = \langle 1,3 \rangle,\langle 2,4 \rangle} 
( c_{l,i,\sigma}^\dagger c_{l+1,j,\sigma} + {\rm h.c.} ) \nonumber \\
&-&  (1-a)t \sum_{l,\sigma} 
\sum_{\langle i,j \rangle = \langle 1,2 \rangle,\langle 3,4 \rangle} 
( c_{l,i,\sigma}^\dagger c_{l,j,\sigma} + {\rm h.c.} )  \nonumber \\
&-& t \sum_{l,\sigma} \sum_{\langle i,j \rangle = {\rm others}}
( c_{l,i,\sigma}^\dagger c_{l,j,\sigma} + {\rm h.c.} ), \\
H_{\rm int} &=& U \sum_{l,i} 
(c_{l,i,\uparrow}^\dagger c_{l,i,\uparrow} - \frac{n_{\rm el}}{2})
(c_{l,i,\downarrow}^\dagger c_{l,i,\downarrow} 
- \frac{n_{\rm el}}{2}) \nonumber \\
&+& \sum_{l,l',i,j} W(r_{l,l',i,j}) 
(\sum_\sigma c_{l,i,\sigma}^\dagger c_{l,i,\sigma} - n_{\rm el})
(\sum_\tau c_{l',j,\tau}^\dagger c_{l',j,\tau} - n_{\rm el}).
\eneqa
In eq. (1), the first term is the tight binding part of 
the $\soc$-polymer backbone, and the second term is the Coulomb 
interaction potential among electrons.  In eq. (2), $t$ 
is the hopping integral between the nearest neighbor carbon 
atoms; $l$ means the $l$th $\soc$ molecule, and $\langle i,j 
\rangle$ indicates the pair of the neighboring $i$th and 
$j$th atoms; the atoms with $i=1 - 4$ of the four-membered 
ring are shown by numbers in Fig. 1, and the other $i$ within 
$5 \leq i \leq 60$ labels the remaining atoms in the same 
molecule; $c_{l,i,\sigma}$ is an annihilation operator of 
the electron at the $i$th site of the $l$th molecule with 
spin $\sigma$; and the sum is taken over the pairs of 
neighboring atoms.  Equation (3) is the Coulomb interactions 
among electrons.  Here, $n_{\rm el}$ is the number of 
electrons per site; $r_{l,l',i,j}$ is the distance between 
the $i$th site of the $l$th $\soc$ and $j$th site of the 
$l'$th $\soc$; and 
\beeq
W(r) = \frac{1}{\sqrt{(1/U)^2 + (r/r_0 V)^2}}
\eneq
is the Ohno potential used in ref. [17].  The quantity 
$W(0) = U$ is the strength of the onsite interaction; 
$V$ means the strength of the long range part; and 
$r_0 = 1.433$\AA~ is the mean bond length of the single 
$\soc$ molecule.  We use the long-range interaction 
because the excited electron and hole spread over a 
fairly large region of the system considered.

The parameter $a$ controls the strength of conjugations 
between neighboring molecules.  This parameter has been 
introduced in the previous papers [9,10].  When $a=1$, 
the $\sigma$-bondings between atoms, 1 and 2, 3 and 4, 
in Fig. 1 are completely broken and the orbitals would 
become like $\pi$-orbitals.  The bonds between the atoms, 
1 and 3, 2 and 4, become double bonds.  As $a$ becomes 
smaller, the conjugation between the neighboring molecules 
decreases, and the $\soc$ molecules become mutually 
independent.  In other words, the interactions between 
molecules become smaller in the intermediate $a$ region.  
In this case, the operator $c_{l,i,\sigma}$ at the lattice 
sites of the four-membered rings represents a molecular 
orbital, in other words, one of the relevant linear 
combinations of the $sp^3$ orbitals.

In the next two sections, we use the system with two $\soc$ 
molecules in order to see CT-exciton components.  The three-molecule
system is considered in the Appendix.  It is of course that 
the much larger systems will be more favorable, but the 
calculations are consumptive of CPU times and huge computer 
memories.  The two molecular system is a minimum system in order 
to see how the CT-components are different between the $\soc$-dimers 
and polymers.  We consider the neutral case and the doped 
case with one electron per $\soc$.  The neutral case 
corresponds to the dimers formed upon UV irradiation and 
to $\soc$-polymers produced under high pressures.  We perform 
the unrestricted Hartree-Fock approximation for operators 
of electrons, and the electron-hole excitation energies are
determined by the single-excitation configuration interaction
method.  We will change the parameter, $a$, within 
$0 \leq a \leq 1.0$.  For Coulomb parameters, we use $U 
= 4t$ and $V = 2t$ from our experiences of the calculations 
on optical excitations in $\soc$ and $\rug$ [13].  The 
same values have been used in the calculations of the 
$\soc$ clusters with neutral charge [14,15].  The energy 
scale of the optical excitations is shown in units of eV 
by using $t=1.8$eV [13].

Figure 1 shows two molecules from the $\soc$-polymer with 
the bond lengths used for the calculations of the polymers.  
The length 9.138\AA~ between the centers of $\soc$ and bond 
lengths, 1.44\AA, 1.90\AA, and 1.51\AA, around the four-membered 
ring are taken from the crystal structure data [6].  The 
other bond lengths, 1.40\AA~ (for the short bonds) and
1.45\AA~ (for the long bonds), are the same as used in the
previous paper [17].   In the calculations of the dimers, 
the bond lengths around the four membered ring between 
the molecules are the same as in Fig. 1, and the other
bond lengths are 1.40\AA~ and 1.45\AA.

The optical spectra of $\soc$ are isotropic with respect 
to the direction of the electric field, but they become 
anisotropic in $\rug$ due to the reduced symmetry [13].
The similar thing occurs in the present calculations.  
In order to simplify calculated data and discussion of 
this paper, we would like to average out over the anisotropy.  
Now we have in mind of the fact that the real $\soc$-dimers 
and polymers are systems where very large orientational 
disorder is present, and we expect that anisotropic 
effects in the optical spectra are not easy to be observed 
experimentally.  The optical absorption is calculated by 
the formula:
\beeq
\sum_\kappa \rho(\omega - E_\kappa) 
(f_{\kappa,x} + f_{\kappa,y} + f_{\kappa,z}),
\eneq
where $\rho(\omega) = \gamma/[\pi(\omega^2+\gamma^2)]$ 
is the Lorentzian distribution of the width $\gamma$,
$E_\kappa$ is the energy of the $\kappa$th optical 
excitation, and $f_{\kappa,x}$ is the oscillator strength 
between the ground state and the $\kappa$th excited state
where the electric field is parallel to the $x$-axis.

\section{CT-EXCITONS IN C$_{\bf 60}$-DIMERS}

Figures 2(a) and (b) show the optical absorption spectra
of the neutral systems, and Figs. 2(c) and (d) show the
absorption of the systems doped with one electron per
$\soc$.  There are three main peak structures which 
are typical to a single molecule in Figs. 2(a) and (b) [13].
In Figs. 2(c) and (d), a small feature appears around
0.4-0.6eV.  This is owing to the filling of electrons
in the $t_{1u}$ orbitals derived from the $\soc$ molecule.
The optical transition from the $t_{1u}$ to the $t_{1g}$
orbitals is dipole allowed, and this gives rise to 
the new peak at the low energies.  The similar fact
has been recently reported in the optical spectra by the 
quantum chemical calculations [18].

In order to characterize the CT-excitons, we calculate
the probability that the excited electron and hole 
exist on the different molecules.  If the probability
is larger than 0.5, we regard the excitation as CT-like.
If it is smaller than 0.5, the excitation tends to 
localize on a single molecule and is Frenkel-like.
We calculate contributions from the CT-excitons to the 
optical absorption, by retaining only the CT-like
excitations in eq. (5).  The results are shown by
the thin lines in Fig. 2.

Even though the molecular exciton character is dominant,
small CT-like components develop in the optical spectra.
In Figs. 2(a) and (b), there are relatively larger 
oscillator strengths of the CT-excitons at the higher
energy sides of the main three features which are
centered around 3.2eV, 4.8eV, and 6.0eV.  In the doped
case of Figs. 2(c) and (d), the absorption of the 
CT-excitons become somewhat broader, but a certain 
amount of CT-like oscillator strengths exist.
We also find that the small features around 0.4-0.6eV,
which appear upon doping, are mainly Frenkel-like.

Now, we summarize the developments of the CT-components
as increasing the conjugation parameter.  The ratio of
the sum of the oscillator strengths of the CT-excitons
with respect to the total absorption is calculated and
is shown as a function of the conjugation parameter in
Fig. 3.  In other words, we have calculated the ratio
of the area below the thin line to the area below the 
bold line in Fig. 2.  The filled and open squares are
the data of the neutral and doped systems, respectively.
We find that the CT-components first increase as the
intermolecular interactions switch on, and saturate
at about $a=0.3$ or so.  It is of some surprise that the
CT-components gradually decrease at the larger $a$ than
about 0.5, i.e., in the large conjugation region. 
Without calculations, one may expect that the value 
will continue increasing as the intermolecular 
interactions become stronger.  We also find that the 
CT-component of the doped system is smaller than that 
of the neutral system.  Thus, the doping effects slightly
suppress the separations between the excited electron
and the hole.

\section{CT-EXCITONS IN C$_{\bf 60}$-POLYMERS}

The calculations are extended to the $\soc$-polymers.
The two molecule system is considered.  The characterization
of the CT-excitons have been done similarly as in the 
previous section. Figures 4(a) and (b) show the optical 
absorption spectra of the neutral system for the two 
conjugation parameters.  The thin line represents the
optical spectra due to the CT-excitons.  The CT-excitons 
exist at the higher energy side of each of the three main 
features.  This property is clearly seen for the stronger
conjugations, $a=0.5$, in Fig. 4(b).  The broadening
of the main three features is larger when the intermolecular 
interactions are stronger.  A small shoulder appears
in the lower energy region than 2.8eV in Fig. 4(b). 
There are dipole-forbidden transitions in these energy
region for the single molecule.  They become partially
allowed when there are intermolecular interactions.
This is the origin of the broad low energy structure.
Such broad structures have been measured in the optical
absorption of $\soc$-polymers synthesized under high
pressures [12], and have been ascribed as the symmetry
lowering effects.

Next, we look at the optical spectra of the doped system
with one electron per $\soc$.  The total absorptions and
the CT-like components are shown for $a=0.2$ and 0.5
in Figs. 4(c) and (d), respectively.  The appearance
of the small feature at 0.4-0.6eV is more distinct than
in the $\soc$-dimers.  As we are treating the two-molecule
system, the small feature is a peak.  But, if we
can do calculations for the infinitely long polymer,
the feature will develop into a Drude tail which is typical 
for metallic systems.  In the present exciton formalism,
we cannot reproduce the Drude structure.  This is the 
limitation of the calculations on small systems.
By comparing the neutral and doped systems, we find that
the CT-components are suppressed upon doping.

The ratio of the oscillator strengths of the CT-excitons
with respect to the total absorption is calculated,
and is shown as a function of the conjugation parameter
in Fig. 5.  The filled and open squares are for the 
neutral and doped systems.  We find that the CT-components
first increase in smaller conjugations, and saturate
at the intermediate conjugations.  The dramatic decrease
is seen in the large conjugations.  The CT-components
of the $\soc$-polymers are larger than those of the 
$\soc$-dimers in Fig. 3.  This is due to the enhanced
intermolecular interactions in polymers.  We also note
that the CT-components decrease when the neutral system
is doped with electrons.  The same qualitative feature
has been seen in the calculations of the dimers.

\section{SUMMARY AND DISCUSSION}

In this paper, we have studied the CT-like excitons in 
optical absorption of the neutral and doped $\soc$-dimers 
and polymers.  We have looked at the variations in 
the optical spectra by changing the conjugation parameter.  
We have found that the total CT-component first increases 
in smaller conjugations, and this is a natural result.
But, it saturates at the intermediate conjugations, and 
the dramatic decrease follows in the large conjugations.
These qualitative features are commonly seen in the 
$\soc$-dimers and polymers, though the strengths of 
CT-like oscillator strengths are different.   We have
also found that the CT-components of the doped systems 
are smaller than those of the neutral systems, and 
thus the electron-hole separations are suppressed 
upon doping in polymerized $\soc$ systems.

It is more favorable to do calculations for much larger
systems.  But, the present exciton formalism over all the
single excitations cannot be applied to systems larger than 
the two molecules.  Then, we have considered the CT-like 
features in the low energies for the three molecule system 
in the Appendix.  The variations of the CT components with 
respect to the conjugation conditions are qualitatively
similar as in the two molecule systems: there is a
saturation of the electron-hole separations at the
intermediate conjugations.  Therefore, we could expect 
that this qualitative point might be seen in larger systems.

\begin{flushleft}
{\bf APPENDIX: THREE MOLECULE SYSTEM OF THE C$_{\bf 60}$-POLYMERS}
\end{flushleft}

\noindent
In the computation circumstances which have been used in the
present paper, it is possible to take into account of all
the single electron-hole pair excitations in the two $\soc$
molecule systems.  When the system size becomes larger, we
cannot treat all of the single excitations.  Alternatively,
we can discuss properties of the low energy excitations by
introducing some cutoffs for the optical excitations with 
high energies.  In this Appendix, we report about the optical 
spectra of the three molecule system of the $\soc$-polymer.  
We consider the optical spectra below the energy about 4eV 
by taking into account of the electron excitations from the 
occupied states ($t_{2u}$, $g_u$, $g_g$, $h_g$, and $h_u$ 
orbitals) to the empty states ($t_{1u}$, $t_{1g}$, $h_g$, 
$t_{2u}$, and $h_u$ orbitals).  Here, the orbitals are in 
the order of the energy values by H\"{u}ckel theory of the 
single $\soc$.

Figure 6 shows the ratio of the oscillator strengths of
CT-excitons with respect to the total absorption calculated
for the neutral three molecules.  There is a maximum at 
about $a= 0.4$, and the CT-component decreases at larger
conjugations.  The maximum value is about 1.6 times as 
large as that of Fig. 3, and it is about 2.2 times as 
large as that of Fig. 5.  This is due to the larger system
size.  However, the property that the CT-component increases
at small conjugations and decreases at larger conjugations
is seen in Fig. 6, too.

\pagebreak
\begin{flushleft}
{\bf REFERENCES}
\end{flushleft}

\noindent
$[1]$ A. M. Rao, P. Zhou, K. A. Wang, G. T. Hager, J. M. Holden,
Y. Wang, W. T. Lee, X. X. Bi, P. C. Eklund, D. S. Cornett,
M. A. Duncan, and I. J. Amster, Science {\bf 259}, 955 (1993).\\
$[2]$ O. B\'{e}thoux, M. N\'{u}\~{n}ez-Regueiro, L. Marques, 
J. L. Hodeau, and M. Perroux, Paper presented at the Materials 
Research Society meeting, Boston, USA, November 29-December 3, 
1993, G2.9.\\
$[3]$ Y. Iwasa, T. Arima, R. M. Fleming, T. Siegrist, O. Zhou,
R. C. Haddon, L. J. Rothberg, K. B. Lyons, H. L. Carter Jr., 
A. F. Hebard, R. Tycko, G. Dabbagh, J. J. Krajewski, G. A. Thomas, 
and T. Yagi, Science {\bf 264}, 1570 (1994).\\
$[4]$ M. N\'{u}\~{n}ez-Regueiro, L. Marques, J. L. Hodeau,
O. B\'{e}thoux, and M. Perroux, Phys. Rev. Lett. {\bf 74}, 278 (1995).\\
$[5]$ O. Chauvet, G. Oszl\`{a}nyi, L. Forr\'{o}, P. W. Stephens,
M. Tegze, G. Faigel, and A. J\`{a}nossy,
Phys. Rev. Lett. {\bf 72}, 2721 (1994).\\
$[6]$ P. W. Stephens, G. Bortel, G. Faigel, M. Tegze,
A. J\`{a}nossy, S. Pekker, G. Oszlanyi, and L. Forr\'{o},
Nature {\bf 370}, 636 (1994).\\
$[7]$ S. Pekker, L. Forr\'{o}, L. Mihaly, and A. J\`{a}nossy,
Solid State Commun. {\bf 90}, 349 (1994).\\
$[8]$ S. Pekker, A. J\`{a}nossy, L. Mihaly, O. Chauvet,
M. Carrard, and L. Forr\'{o}, Science {\bf 265}, 1077 (1994).\\
$[9]$ K. Harigaya, Phys. Rev. B {\bf 52}, 7968 (1995).\\
$[10]$ K. Harigaya, Chem. Phys. Lett. {\bf 253}, 420 (1996).\\
$[11]$ M. S. Dresselhaus, G. Dresselhaus, and P. C. Eklund,
{\sl Science of Fullerenes and Carbon Nanotubes}
(Academic Press, San Diego, 1996).\\
$[12]$ Y. Iwasa, {\sl Optical Properties of Low-Dimensional
Materials}, edited by T. Ogawa and Y. Kanemitsu 
(World Scientific, Singapore, 1995) Chap. 7.\\
$[13]$ K. Harigaya and S. Abe, Phys. Rev. B {\bf 49}, 16746 (1994).\\
$[14]$ K. Harigaya and S. Abe, Mol. Cryst. Liq. Cryst. {\bf 256},
825 (1994).\\
$[15]$ K. Harigaya and S. Abe, {\sl 22nd International Conference
on the Physics of Semiconductors} (World Scientific, Singapore,
1995), p. 2101.\\
$[16]$ K. Harigaya, Phys. Rev. B {\bf 50}, 17606 (1994).\\
$[17]$ K. Harigaya, Phys. Rev. B {\bf 53}, R4197 (1996).\\
$[18]$ J. Fagerstr\"{o}m and S. Stafstr\"{o}m, Phys. Rev. B {\bf 53},
(1996) (in press).\\

\pagebreak

\begin{flushleft}
{\bf FIGURE CAPTIONS}
\end{flushleft}

\mbox{}

\noindent
Fig. 1.  The structures of two molecule system of the 
$\soc$-polymer.  The carbon sites which constitute the
four membered rings are named with numbers.  The bond 
lengths, used in the calculations, are also shown.

\mbox{}

\noindent
Fig. 2.  The optical absorption spectra of $\soc$-dimers
for the neutral case with (a) $a = 0.2$ and (b) $a = 0.5$,
and for the one-electron (per $\soc$) doped case with (c) 
$a = 0.2$ and (d) $a = 0.5$.  The bold line represents the 
total absorption, and the thin line shows the contribution
from the charge-transfer excitons.  The broadening
$\gamma = 0.108$eV is used.

\mbox{}

\noindent
Fig. 3.  The ratio of the oscillator strengths 
(CT-component) of the charge-transfer excitons with 
respect to the total oscillator strengths in the 
$\soc$-dimers.  The closed squares are for the neutral 
cases and the open squares are for the one-electon (per $\soc$)
doped cases.

\mbox{}

\noindent
Fig. 4.  The optical absorption spectra of $\soc$-polymers
for the neutral case with (a) $a = 0.2$ and (b) $a = 0.5$,
and for the one-electron (per $\soc$) doped case with (c) 
$a = 0.2$ and (d) $a = 0.5$.  The bold line represents the 
total absorption, and the thin line shows the contribution
from the charge-transfer excitons.  The broadening
$\gamma = 0.108$eV is used.

\mbox{}

\noindent
Fig. 5. The ratio of the oscillator strengths 
(CT-component) of the charge-transfer excitons with 
respect to the total oscillator strengths in the 
$\soc$-polymers.  The closed squares are for the neutral 
cases and the open squares are for the one-electon 
(per $\soc$) doped cases.

\mbox{}

\noindent
Fig. 6. The ratio of the oscillator strengths 
(CT-component) of the charge-transfer excitons with 
respect to the total oscillator strengths in the 
neutral three molecule system.

\end{document}